\documentclass[aps,prl,reprint,groupedaddress]{revtex4-2}

\usepackage{graphicx}
\usepackage{amsmath,amssymb,multirow}
\usepackage{bm}
\usepackage{braket}
\usepackage[
colorlinks,
linkcolor=blue,
urlcolor=blue,
citecolor=blue
]{hyperref}

\begin{document}

\title{Direct laser cooling of Rydberg atoms with an isolated-core transition}

\author{A.\,Bouillon}
\author{E.\,Marin-Bujedo}
\author{M.\,G\'en\'evriez}
\email[]{matthieu.genevriez@uclouvain.be}
\affiliation{Institute of Condensed Matter and Nanosciences, Universit\'e catholique de Louvain, BE-1348 Louvain-la-Neuve, Belgium}

\date{\today}

\begin{abstract}
We propose a scheme to directly laser cool Rydberg atoms by laser cooling the residual ion core within the Rydberg-electron orbit. The scheme is detailed for alkaline-earth-metal Rydberg atoms, whose ions can be easily laser cooled. We demonstrate that a closed optical cooling cycle can be found despite the perturbations caused by the Rydberg electron, and that this cycle can be driven over more than $100~\mu$s to achieve laser cooling. The cooling dynamics with and without the presence of magnetic fields are discussed in detail.
\end{abstract}

\maketitle

Since atoms and ions were first laser cooled, opening a fascinating window into
ultracold matter, sustained effort has been made to laser cool increasingly
complex systems. Recent years witnessed the laser cooling of
diatomic~\cite{shuman10,zhelyazkova14} and small
polyatomic~\citep{kozyryev17,mitra20} molecules, motivated by their interest in
quantum information~\cite{demille02} and simulation~\cite{blackmore18} or in
testing the standard model~\citep{augenbraun20}. Rydberg atoms are another
important class of complex systems, whose large density of states and extreme
sensitivity to their environment~\cite{gallagher94} are a key feature of many
applications in quantum simulation~\citep{browaeys20,whitlock17,nguyen18},
information~\cite{cohen21} and optics~\cite{adams20}. Rydberg atoms are
routinely prepared at ultracold temperatures by photoexcitation of ground state
atoms in optical traps~\cite{walz-flannigan04,cote06,macri14,endres16}, and
trapped with electric fields~\citep{hogan08}, magnetic fields~\cite{choi05}, or
optical fields~\citep{mukherjee11,cortinas20}. Divalent Rydberg atoms have also
been trapped in tightly focused optical tweezers~\cite{wilson22,holzl24}.
However, Rydberg atoms have never been directly laser cooled. Doing so so would
open new ways to cool or manipulate ensembles of strongly interacting Rydberg
atoms at low temperatures, and pave the way to, e.g., combining laser cooling
with tuneable long-range interactions~\cite{glaetzle12}, exploring the
thermodynamics of strongly interacting Rydberg
gases~\cite{osychenko11,brune20}, or dissipating the heat generated when
Rydberg atoms are used to sympathetically cool polar
molecules~\cite{zhang23,zhao12}.

In general the dense energy-level structure associated with the Rydberg
electron is not suitable for Doppler cooling because no closed cooling cycle is
available. The only exception is between adjacent circular Rydberg states
($l=n-1$, $|m_l|=l$), but even in this case the low photon energy associated
with the transition and the long fluorescence lifetime would make cooling
prohibitively slow. To overcome this difficulty, two schemes have been proposed
that use atoms prepared in a coherent superposition of low-lying electronic
states, which are used for laser cooling, and a Rydberg
state~\citep{guo96,bounds18}. In the only existing experimental work, Sr atoms
with a small Rydberg character (1\%) admixed to low-lying excited states were
for example laser cooled to $\sim 1\ \mu$K~\cite{bounds18}. The laser cooling
of Rydberg atoms themselves, without laser dressing, is yet to be achieved.

\begin{figure}
	\includegraphics[width=\columnwidth]{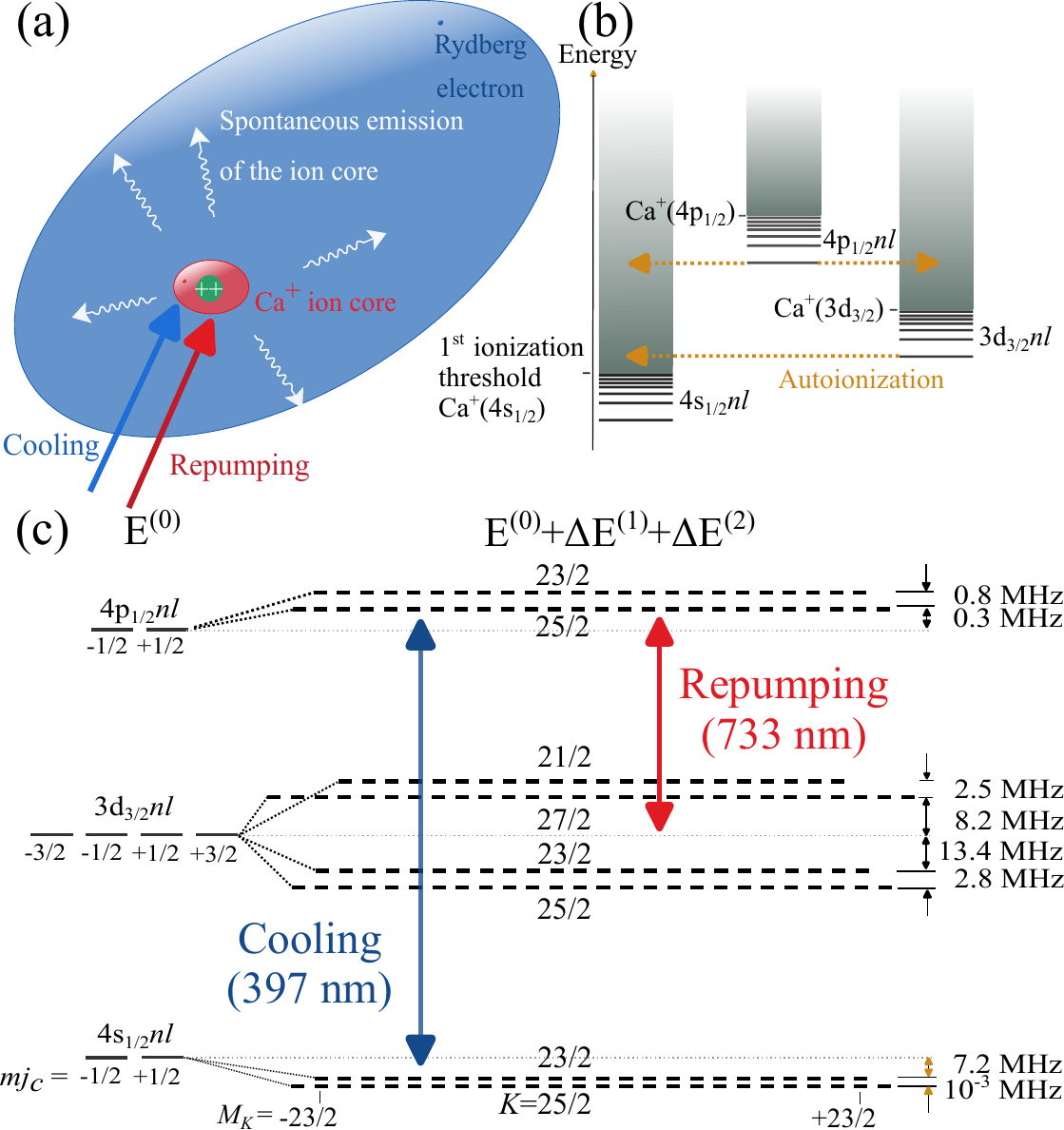}
	\caption{(a) Schematic representation of Rydberg-atom laser cooling with an isolated-core transition. (b) Schematic view of the energy-level structure of $4s_{1/2}nl$, $4p_{1/2}nl$ and $3d_{3/2}nl$ Rydberg series. (c) Energy-level structure of the relevant $^{40}$Ca Rydberg states ($n=55, l=12$), without ($E^{0}$) and with ($E^{0}+\Delta E^{(1)}+\Delta E^{(2)}$) residual electron interaction. Energy splittings are in units of $h\cdot$MHz.\label{fig:scheme}}
\end{figure}

In this letter, we propose and theoretically analyze a scheme for directly
laser cooling Rydberg atoms. It is based on the premises that, in an atom or a
molecule excited to a high Rydberg state, the residual ion core is essentially
\emph{isolated} from the Rydberg electron and can be excited and optically
manipulated as if it were an isolated
ion~\cite{cooke78,wehrli19,genevriez21b,muni22}. It is then conceivable, at
least in principle, to laser cool the ion core within the Rydberg-electron
orbit, thereby cooling the Rydberg atom itself [see Fig.~\ref{fig:scheme}(a)].
We demonstrate below that, for Rydberg states of alkaline-earth atoms with high
angular momentum, a strong resonant-radiation-pressure force can be applied
over a timescale longer than the typical $>$ 100 $\mu$s radiative lifetime of
the Rydberg electron. Furthermore, this force is applied without significantly
perturbing the Rydberg electron, an interesting feature for many applications
of Rydberg atoms~\cite{saffman10,browaeys20}.

Alkaline-earth atoms are ideal systems for Rydberg-atom laser cooling because
the electronic structure of their ion core is, as for isolated
ions~\cite{eschner03}, simple and well suited for Doppler cooling. We consider
below the case of $^{40}$Ca for illustration, however the proposed scheme can
be extended to other isotopes, other alkaline-earth species
and, in principle, any other element of the periodic table whose ion can be
laser cooled. The Doppler laser cooling of $^{40}$Ca$^+$ involves the
$4s_{1/2} - 4p_{1/2}$ transition ($A=1.4~10^8$~s$^{-1}$) and the $3d_{3/2} -
4p_{1/2}$ repumping transition to close the cooling cycle
[Fig.~\ref{fig:scheme}(c)]. Our scheme for Rydberg atoms uses the same ion-core
states. However, the residual Coulomb repulsion between the Rydberg electron,
labeled by its principal-, orbital-angular-momentum- and magnetic quantum
numbers $(n, l, m_l)$, and the electrons of the ion core makes the cooling
dynamics more complex.

First, the doubly excited states we consider here ($4p_{1/2}nl, 3d_{3/2}nl$) are above the first ionization threshold [Fig.~\ref{fig:scheme}(b)].
Because of the electronic repulsion the atom may rapidly autoionize.
Autoionization rates scale as $n^{-3}$, which means that autoionization could
be suppressed by increasing $n$. This approach is however impractical because,
for $l$ values close to unity, $n$ values of the order of $1000$ would be
required to reach lifetimes long enough to carry out a sufficient number of
cooling cycles ($\sim 10^4$)~\cite{fields18}. Rydberg states with such
principal quantum numbers are challenging to produce experimentally
and exceedingly sensitive to external perturbations~\cite{ye13}. Alternatively,
autoionization can be suppressed by increasing $l$. We recently showed that the
rates decay exponentially with $l$ and that, for the typical values of $n \sim
50$ used in cold Rydberg-atoms experiments, the autoionization lifetime is
comparable to the Rydberg-electron radiative lifetime already for $l \gtrsim
10$~\cite{marin-bujedo23}. Experimentally, states with $l\sim 10$ can be
prepared using Stark switching~\citep{lehec21} whereas circular states are
prepared with microwave fields or crossed electric and magnetic
fields~\citep{morgan18,hulet83a}. We consider below Ca Rydberg states with $l =
12$, the value at which the autoionization rates of all relevant states become
smaller than those of the radiative decay of the Rydberg
electron~\cite{marin-bujedo23}. States with lower $n$ and $l$ values, which
are more amenable to Stark switching, can also be used as discussed below.

The second consequence of the residual electronic correlation between the
Rydberg- and ion-core electrons is the increase of the number of energy levels
relevant for cooling compared to the isolated ion, and the shift of their energies [Fig.~\ref{fig:scheme}(c)]. We detail their calculations below.

In a high-$l$ state the large centrifugal barrier $l(l+1)/2r^2$ that the
Rydberg electron experiences prevents its penetration in the ion-core region.
The Rydberg electron experiences an almost purely Coulombic potential, and its
wavefunction is thus well described by a hydrogenic orbital. Because the
distance between the core- and Rydberg electrons is always large, the Coulomb
repulsion between them is weak and well treated within first- and second-order
perturbation theory~\cite{chen92,shuman06,marin-bujedo23}. Moreover, the
spin-orbit interaction of the Rydberg electron, scaling as
$n^{-3}$~\citep{gallagher94}, is negligible compared to, e.g., the natural
linewidths of the cooling and repumping transitions. Rydberg states are
well described within the $jK$ coupling scheme~\citep{aymar96}, in
which the total electronic angular momentum of the ion core ($\bm{j}_c$)
couples to the orbital angular momentum of the Rydberg electron ($\bm{l}$) to
give $\bm{K}$.

A zeroth-order electronic wavefunction, neglecting residual electron
correlations between the Rydberg electron and the ion core, is given by
\begin{equation}
\begin{split}
	\ket{n_cl_cj_cnlKM_Ksm_{s}} =& \sum_{m_{j_c},m_{l}} \braket{j_cm_{j_c}lm_{l}|KM_K} \\
          &\ket{n_cl_cj_cm_{j_c}}\ket{nlm_{l}}\ket{sm_{s}} ,
\end{split}
	\label{eq:jkcoupling}
\end{equation}
where $\ket{n_cl_cj_cm_{j_c}}$ describes the electronic wavefunction of the ion core and
$\ket{nlm_{l}}$ is the Rydberg-electron orbital. We can already note that,
because of angular-momentum coupling, the two degenerate magnetic sublevels
that exist for Ca$^+(4s_{1/2})$ have now turned into two sets of 24 and 26
degenerate $M_K$ sublevels associated with the $K=23/2$ and $25/2$ levels,
respectively [Fig.~\ref{fig:scheme}(c)].

The energy of the Rydberg states, neglecting residual electron correlations, is given by the Rydberg formula
\begin{equation}
	E^{(0)} = E_{n_cl_cj_c}^{\text{Ca}^+} - \frac{Ry}{n^2} ,
\end{equation}
where $Ry$ is the Rydberg constant for $^{40}$Ca. The first-order correction $\Delta E^{(1)}$ to this energy corresponds to the interaction between the permanent quadrupole moment of the ion core $\Theta(n_c,l_c,j_c)$ and the electric field generated by the outer electron~\cite{chen92},
\begin{align}
	\Delta E^{(1)} =& (-1)^{K+j_c} \braket{nl|r^{-3}|nl} (2l+1)\Theta(n_c,l_c,j_c)  \nonumber\\
	&\times \begin{pmatrix}
		l & 2 & l \\
		0 & 0 & 0
	\end{pmatrix}
	\begin{Bmatrix}
		K & l & j_c \\
		2 & j_c & l
	\end{Bmatrix}{\bigg /}\begin{pmatrix}
		j_c & 2 & j_c \\
		j_c & 0 & -j_c
	\end{pmatrix}.
\end{align}
Because of symmetry, it is nonvanishing for the $3d_{3/2}$ ion-core state only,
which explains the greater energy shifts of the associated Rydberg states
compared to the $4s_{1/2}$ and $4p_{1/2}$ ion-core states. The second-order
energy correction $\Delta E^{(2)}$ corresponds to the polarization of the ion
core by the outer electron. Its expression is more complex and given in
\cite{shuman06}.

We calculated the first- and second-order energy corrections using known
multipole moments, matrix elements and polarizabilities of
Ca$^+$~\cite{safronova11,singh18,kaur21}. The resulting energy-level structure
of the Ca Rydberg levels involved in the cooling cycle is shown in
Fig.~\ref{fig:scheme}(c) for $n=55$ and $l=12$. Each ion-core state is shifted
and split into several $K$ levels with different energies. Whereas the energy
splittings between the different $K$ states of the $4s_{1/2}$ and $4p_{1/2}$
core levels are small ($< 1~$MHz), and in fact smaller than the natural
linewidth of the $4s_{1/2}-4p_{1/2}$ cooling transition and the
$3d_{3/2}-4p_{1/2}$ repumping transition (24 MHz), the splittings for the
$3d_{3/2}$ core level are larger (up to $26.9~$MHz for the outermost $K$
states).

To check the validity and accuracy of the perturbative treatment presented
above, we compared the energy shifts $\Delta E^{(1)} + \Delta E^{(2)}$ against
those we obtained from nonperturbative \textit{ab initio} calculations using
the method of configuration interaction with exterior complex scaling
(CI-ECS)~\citep{genevriez19,genevriez21}. Briefly, CI-ECS treats the motion of
the two valence electrons of Ca explicitely and without approximation whereas
the effect of the closed shell Ca$^{2+}$ core is accounted for with a model
potential~\cite{genevriez21}. The large two-electron Hamiltonian is calculated
and diagonalized in a basis of two-electron functions built from numerical
basis functions~\cite{rescigno00} and using exterior-complex scaling to
describe continuum processes such as autoionization. Details on the CI-ECS
calculations can be found in Ref.~\cite{marin-bujedo23}. The differences
between the perturbative and CI-ECS energies for the $4s_{1/2}nl$, $4p_{1/2}nl$
and $3d_{3/2}nl$ Rydberg states are very small ($< 1$ MHz), which validates the
perturbative approach. The CI-ECS results also confirm that channel
interactions, which are not accounted for in the single-configuration
perturbative model described above, are negligible for the states under
consideration. Overall, CI-ECS calculations are much more computationally heavy
than perturbative ones. This precludes their use to simulate the time-dependent
photoexcitation and cooling dynamics, to which we now turn.

When light is close to resonance with a transition in the isolated ion, Rydberg
atoms are known to undergo isolated core excitation (ICE)~\cite{cooke78}. The
ion core, which is to a good approximation isolated from the Rydberg electron,
is resonantly excited as if it were an isolated ion while the Rydberg electron,
whose interaction with the radiation is negligible in comparison, is a
spectator of the core excitation (see~\cite{lehec21,yoshida23} for recent
examples). The electric dipole moment of the $4s_{1/2}nl_K -
4p_{1/2}n'l'_{K'}$ transition can be written as~\cite{bhatti81},
\begin{align}
	&\braket{4p_{1/2}n'l'  K'M_K'| \hat{\bm{D}} | 4s_{1/2}nl K M_K} 
    \simeq \nonumber (-1)^{2K-M'_K-l-j'_c}\\& [K, K']^{1/2} \overline{\mu}_c \braket{n'l'|nl}
	\begin{pmatrix}
		K & 1 & K' \\
		M_K & q & -M'_K
	\end{pmatrix}
	\begin{Bmatrix}
		K' & 1 & K \\
		1/2 & l & 1/2
	\end{Bmatrix},\label{eq:ice}
\end{align}
where the right-hand-side is obtained using Eq.~\eqref{eq:jkcoupling} and
assuming that the transition dipole moment for the Rydberg electron
$\braket{n'l' | \bm{\epsilon}\cdot\bm{r} | nl}$ is vanishingly small. We used the standard notation $[K]=2K+1$. The
Wigner $3j$ and $6j$ symbols result from angular momentum coupling, and
$\overline{\mu}_c$ is the reduced dipole moment of the
Ca$^+(4s_{1/2}-4p_{1/2})$ transition. $\braket{n'l'|nl}$ is the overlap
integral between the initial- and final Rydberg-electron wavefunction. Because
the Rydberg electron is quasi hydrogenic, the overlap integral is 1 if $n = n'$
and $l' = l$, and 0 otherwise. Importantly, this shows that the state of the
Rydberg electron is left unchanged upon ICE. Equation~\eqref{eq:ice} also
yields the ICE selection rules $\Delta K = 0, \pm 1$ and $\Delta M_K = q$,
where $q=0, \pm 1$ for $\pi$ and $\sigma^\pm$ polarizations, respectively. A similar expression is obtained for the repumping transition.

Population dynamics when driving the cooling and repumping ICE transitions were
calculated by numerically solving the Lindblad master equation
\cite{johansson12} for the 200-level system shown in Fig~\ref{fig:scheme}. Rabi
frequencies and fluorescence rates obtained from Eq.~\eqref{eq:ice} were
included together with the energies calculated with the perturbative approach
and the autoionization rates reported in Ref.~\cite{marin-bujedo23}. The
latter rates ($<10^3$~s$^{-1}$) do not produce observable population losses
over the time scale of the calculations ($100$~$\mu$s), whose duration was
chosen to correspond to the blackbody-radiation lifetime~\cite{gallagher94} of
a $n=55, l=12$ Rydberg electron.

\begin{figure}
	\centering
	\includegraphics[width=\columnwidth]{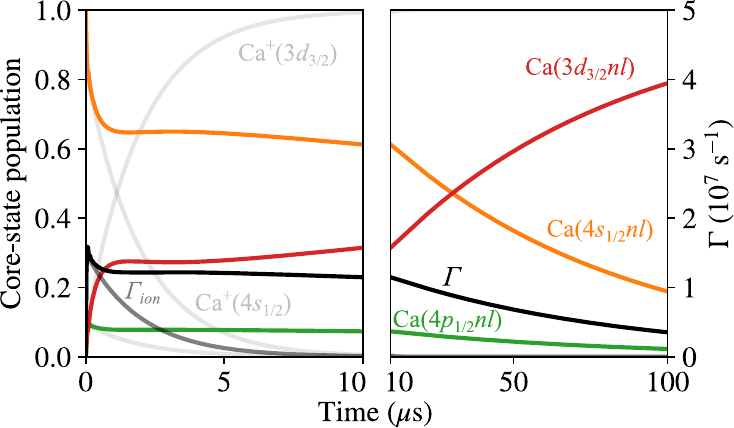}
	\caption{Time evolution of the populations of Rydberg states ($n=55, l=12$) with $4s_{1/2}$ (orange), $3d_{3/2}$ (red) and $4p_{1/2}$ (green) ion core states, obtained by summing the populations of states with different values of $K$ and $M_K$ but identical values of $n_c, l_c$ and $j_c$. The photon-scattering rate $\Gamma$ for the cooling transition is shown as the black line. The dynamics for the isolated ion are shown by the pale gray lines.}
	\label{fig:pops_noBfield}
\end{figure}

Laser parameters for the calculation were chosen based on a detailed study
performed to optimize the cooling of trapped alkaline-earth
ions~\cite{lindvall12}. The cooling-laser intensity is equal to the saturation
intensity of the Ca$^+(4s_{1/2}-4p_{1/2})$ transition ($I_s =
4.66\cdot10^{-4}$~Wmm$^{-2}$) and the detuning relative to that transition is
$\Delta\omega_c/2\pi = 7.7$~MHz, a value that compensates the energy-level
shifts $\Delta E^{(2)}$ (see Fig.~\ref{fig:scheme}). The repumping-laser
intensity is $I_r = I_s$ and its detuning relative to the
Ca$^+(3d_{3/2}-4p_{1/2})$ transition is $\Delta\omega_c/2\pi = 3.2$~MHz. We
assume that the Rydberg atom is initially prepared in the
$4s_{1/2}55(l=12)_{K=23/2}$ state with $M_K=1/2$, which corresponds to the
typical states populated in Stark-switching experiments \cite{freeman76}. We
also verified that starting from another $K$ and $M_K$ state, as would be the
case if, for example, a circular Rydberg state were
considered~\cite{teixeira20}, does not change the overall dynamics nor the
conclusions drawn below.

Figure~\ref{fig:pops_noBfield} shows the population dynamics and photon
scattering rate $\Gamma$ when both lasers have $\pi$ polarizations. It is well
known that, in the isolated ion and for $\pi$ polarizations, the population is
rapidly pumped into the $3d_{3/2}(|m_j| = 3/2)$ dark states and the
photon-scattering rate drops to zero (light gray lines in
Fig.~\ref{fig:pops_noBfield})~\cite{lindvall12}. The same dynamics occur in the
Rydberg atom but over a much longer timescale, with a dark-state pumping rate
determined to be $\Gamma_\text{DS}=1.3\cdot10^4$~s$^{-1}$. The reason for this
lower rate is that the population of the states with an excited $4p_{1/2}$ ion
core is now distributed over the 50 different $M_K$ sublevels ($K=23/2, 25/2$),
and only two of those sublevels can fluoresce to the $3d_{3/2}55(l=12)_{27/2}$, $|M_K| = 27/2$ dark states (see Fig.~\ref{fig:scheme}). When the lasers
have $\sigma^+$ or $\sigma^-$ polarizations, the system decays much faster into
a dark state.
\begin{figure}
	\centering
	\includegraphics[width=\columnwidth]{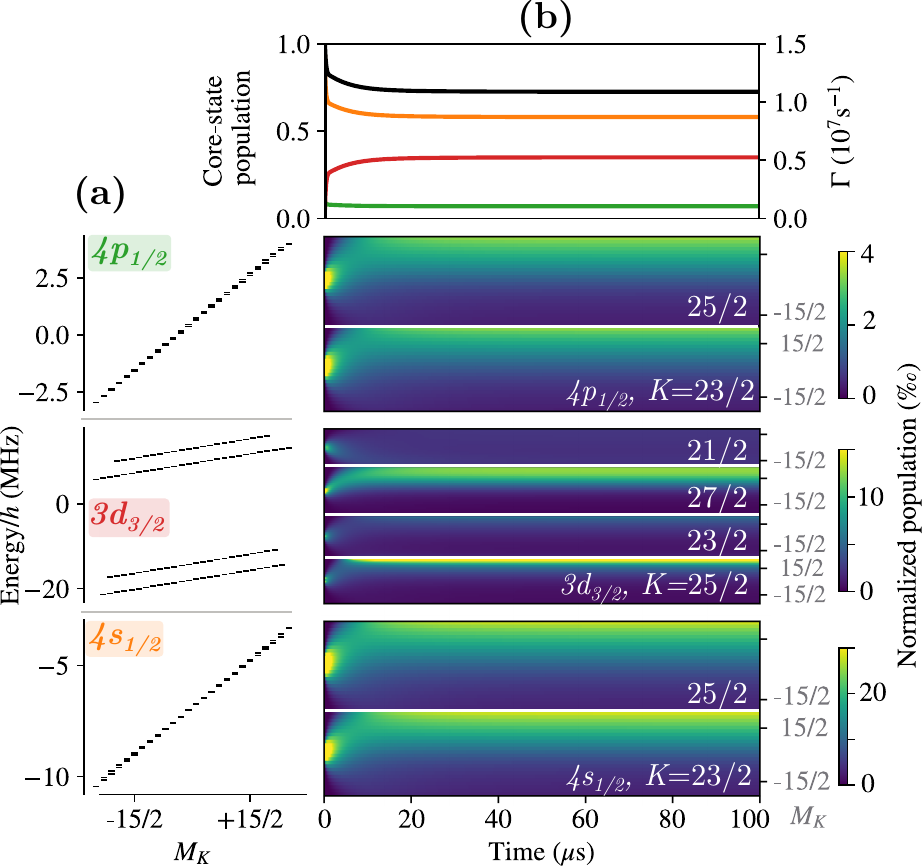}
	\caption{(a) Energy levels of the $4s_{1/2}nl_{K M_K}$,
	$3d_{3/2}nl_{K M_K}$ and $4p_{1/2}nl_{K M_K}$ states in the
	presence of a $20~\mu$T magnetic field ($n=55, l=12)$. (b) Top graph: time evolution of the ion-core
	populations and photon-scattering rate (same as Fig.~\ref{fig:pops_noBfield}). Lower graphs: time evolution of the population of each of the 200 sublevels involved in the cooling. }
	\label{fig:optical_pumping}
\end{figure}

The results of Fig.~\ref{fig:pops_noBfield} reveal that light can be scattered
by the ion core of a Rydberg atom at a rate of up to $\sim 10^7$~s$^{-1}$, a
value similar to the one of the isolated ion. Such a rate translates into a
resonant-radiation-pressure force of $2 \cdot 10^{-20}$~N and a
deceleration of the $^{40}$Ca Rydberg atoms of $3\cdot 10^4~g$. Even in
the absence of a strategy to destabilize dark states, photon scattering is
efficient beyond the $10$-$\mu$s range, a timescale over which the atoms can
be, \textit{e.g.}, cooled or deflected in optical molasses.

To achieve laser-cooling durations over the entire radiative lifetime of the
Rydberg electron, the dark states must be destabilized by, \textit{e.g.},
periodically switching the laser polarization or applying a magnetic
field~\cite{berkeland02,lindvall12}. We consider below the  magnetic-field
case, which further offers the opportunity to assess the effect of stray
magnetic fields present in experiments. For a magnetic field $B$ of moderate
strength, the ion core undergoes a Zeeman effect in the weak-field limit
whereas for the Rydberg electron it is well into the Paschen-Back
regime~\cite{gallagher94} because the spin-orbit interaction is very small. The
Zeeman Hamiltonian is then diagonal in the $\ket{K M_K}$ basis with eigenvalues
given by
\begin{align}
	\Delta E^{\text{Zeeman}}_{K, M_K} = \mu_B B\sum_{m_{j_c}, m_{l}} & {\braket{j_cm_{j_c}lm_{l}|KM_K}}^2 
	\\ &\times\left[g_{j_c}m_{j_c} + m_{l} + g_s m_{s}\right],\nonumber
\end{align}
where $g_{j_c}$ and $g_s$ are the ion-core and Rydberg-electron $g$ factors,
respectively. The energy level structure of the Zeeman-split levels is shown in
Fig.~\ref{fig:optical_pumping} (a) for a magnetic field $B=20~\mu$T. The
splitting between two adjacent $M_K$ levels is $\sim0.3$~MHz, a value
sufficient to destabilize the dark states at a rate larger than
$\Gamma_{\text{DS}}$.

Population dynamics calculated for a $\pi$-polarized cooling laser and a
repumping laser that is elliptically polarized (90\% of the intensity in
$\sigma^+$ and 10\% in $\sigma^-$), are shown in
Fig.~\ref{fig:optical_pumping}(b). These polarizations were chosen to
illustrate two effects. First, the magnetic field effectively removes the dark
states and a photon-scattering rate of $\sim 10^7$ s$^{-1}$ can be maintained
beyond 100~$\mu$s. A field of $20~\mu$T is weaker than the typical Earth
magnetic field and therefore, unless residual magnetic fields are actively
compensated in an experiment, they will destabilize dark states and allow
Rydberg-atom laser cooling over long timescales. Second, by choosing an
appropriate polarization for the repumping laser, the atoms can be optically
pumped into, predominantly, the largest $M_K$ sublevels [see
Fig.~\ref{fig:optical_pumping}(b)]. Other ellipticities for the repumping-laser
polarization influence slightly the degree of optical pumping and the
photon-scattering rate, however a large $\Gamma$ can still be maintained beyond
100~$\mu$s as well.

Large photon scattering rates can also be achieved for stronger magnetic
fields. For example, for $B=6$~mT, the Zeeman shifts of the $\Delta M_K=\pm1$
transitions (approximately $\pm 84$~MHz) can be compensated by tuning the
frequencies of the $\sigma^+$ and $\sigma^-$ components of the cooling and
repumping lasers accordingly, reaching $\Gamma > 5\cdot 10^{6}$~s$^{-1}$.
Interestingly, the tuneability of the cooling-transition frequency with an
external magnetic field makes it possible to address Doppler shifts and laser
cool relatively warm Rydberg atoms, much as in conventional Zeeman slowers and
magneto-optical traps. Velocities in the $35$-m/s range ($E/k_\text{B}\sim
3$~K) could be addressed with $B=6$~mT, which corresponds to typical velocities
of atoms exiting a Zeeman slower or a buffer-gas cooling cell~\cite{hutzler12}.
Such atom sources are thus ideal for future experimental implementations of
Rydberg-atom laser cooling. Larger velocities and temperatures can also be
targeted to optically deflect or manipulate Rydberg atoms, within the
limitations imposed by the Stark switching process and the radiative lifetime
of the Rydberg electron.

For $n\sim 50$, we estimate that Rydberg-atom densities larger than $N_0 \sim
10^{10}$~cm$^{-3}$ can be reached. At such densities the Rydberg atoms are
separated by $5$~$\mu$m on average. Long-range dipole-dipole interactions
shift the energy levels by $\sim 15$~MHz~\cite{gallagher94}, a value
comparable to the width of the cooling transition. However, the cooling transition can shift less than the levels themselves, therefore $N_0$ represents a lower
bound of the achievable atomic density. A better estimate of
$N_0$ can only be obtained from accurate calculations of long-range
interactions, which are currently unavailable.

We finally discuss the range of $n$ and $l$ values for which the method is
applicable. The population in the $4p_{1/2}nl_K$ excited states is $\sim 0.1$,
and therefore the rate at which the Rydberg atoms autoionize under laser
cooling is one order of magnitude lower than the one of pure $4p_{1/2}nl_K$
states. This argument holds because the $3d_{3/2}nl_{K}$ states autoionize at
a rate two orders of magnitude smaller than $4p_{1/2}nl_K$
ones~\cite{marin-bujedo23}. Taking the autoionization rates calculated
in~\cite{marin-bujedo23}, the present Rydberg-atom laser cooling scheme is
applicable, for $^{40}$Ca, to $n>40$ for $l=10$, and $n\geq 13$ for $l\geq
11$. In the Sr atom, the larger autoionization rates lead to the cooling
scheme being applicable to $n\geq 32$ for $l = 11$ and $n\geq 13$ for $l \geq
12$. Such states (e.g., $n=17, l=12$) have been populated by Stark switching
in earlier experiments~\cite{eichmann03}.

In conclusion, we have presented a strategy to directly and selectively cool
Rydberg atoms using isolated-core excitation. It requires high $l$ ($l
\gtrsim 10$) Rydberg states to prevent the atoms from autoionizing as they are
cooled. Efficient cooling, with a photon-scattering rate of $~ 10^7$~s$^{-1}$,
can be maintained over more than 100~$\mu$s, i.e., over the radiative
lifetime of the Rydberg electron. Because the Rydberg electron is essentially
hydrogenic, the details of the ion-core structure are expected to have little
influence on the cooling strategy presented above, which should thus be
applicable to isotopes and species other than $^{40}$Ca. The present work paves
the way to an experimental demonstration of direct and selective Rydberg-atom
laser cooling, a result of importance for quantum simulation and metrology. It provides a way to produce ultracold clouds of \emph{only} Rydberg atoms
and to control their temperature, thus permitting the exploration of the
thermodynamical properties of strongly interacting Rydberg gases as a function
of temperature.

\begin{acknowledgments}
This work was supported by the Fonds de la Recherche Scientifique - FNRS under
MIS Grant No. F.4027.24 and IISN Grant No. 4.4504.10. E.M.B. is a FRIA grantee
of the Fonds de la Recherche Scientifique - FNRS. E.M.B. and M.G. acknowledge
support from the Fonds Spéciaux de Recherche (FSR) of UCLouvain.
\end{acknowledgments}

\bibliographystyle{apsrev4-2}

\end{document}